\documentclass{ichep04ask}

\begin{document}

\title{Search For Extra Dimensions At LEP}

\author{S. Ask}

\address{Department of Physics, Lund University, S$\ddot{o}$lvegatan 14,\\
SE-223 63 Lund, Sweden.\\ E-mail: stefan.ask@hep.lu.se}  

\twocolumn[\maketitle\abstract{
Searches for extra dimensions have been made at the four LEP experiments,
ALEPH, DELPHI, L3 and OPAL, where different processes have been searched for within 
both the ADD and Randall-Sundrum scenarios. Since no signs of any
signal have been observed the results have been used to set exclusion
limits in the corresponding extra dimension models. This talk presents 
both individual results from the LEP experiments as well as combined 
results from the searches for graviton emission and virtual graviton exchange. 
The results are mainly based on the data recorded between the years 1998-2000, 
which corresponds to an integrated luminosity of 0.6 fb$^{-1}$ per experiment and 
center-of-mass energies from 189 up to 209 GeV.}]

\section{Introduction}

The data recorded by the four LEP experiments, ALEPH, DELPHI, L3 and OPAL have been 
used to search for extra spatial dimensions. Within these searches effects from both
graviton emission\cite{gecomb} and virtual graviton exchange\cite{vgcomb} have been 
studied together with effects from TeV strings\cite{l3cite}, branons\cite{l3cite} and 
radions\cite{opalcite}. The different searches have mainly been based on the data 
recorded between the year 1998 and 2000, when the LEP machine was running at 
center-of-mass energies from 189 up to 209 GeV.The integrated luminosity was 0.6 fb$^{-1}$ 
per experiment. 

Two different extra dimension scenarios have been studied by the LEP experiments, the so called
ADD and Randall-Sundrum scenarios. All the Standard Model (SM) particles are assumed in both 
scenarios to be located on a brane with three spatial dimensions while gravity is
allowed to propagate in a higher dimensional space, normally called the bulk. This brane picture
is suggested by different string models and the reason why gravity appears to be weak at the SM
brane is due to its access to the more extended space which the SM particles do not have. This
could then give rise to gravitational phenomena at the TeV scale which would solve the so called 
hierarchy problem. 

Gravity propagating in extra dimensions would also give rise to specific phenomena due to a number 
of new particles observed on the SM brane. From macroscopical observations it is clear that the 
extra dimensions would have to be compactified. Their finite size together with the fact that momenta in
the extra dimensions would appear as mass in the 4-dimensional world imply that gravitons have to
appear in so called Kaluza-Klein (KK) towers, where each mass mode corresponds to the total momentum 
``hidden'' in the extra dimensions. The effective 4-dimensional theory of gravity valid at the
energies of LEP ($\sqrt{s}$ lower than a new fundamental scale of gravity $M_{D}$)
then contains three different kinds of new particles which would interact with the SM fields:
\begin{itemize}
\vspace*{-0.1cm} 
\item{massive spin-2 KK gravitons ($G^{(n)}_{\mu \nu }$});
\vspace*{-0.2cm} 
\item{massive scalar KK gravitons ($H^{(n)}$});
\vspace*{-0.2cm} 
\item{massive scalar branons ($\tilde{\pi }$});
\vspace*{-0.1cm} 
\end{itemize}
where the scalar gravitons include the so called radion ($n=0$) and the so called branons are related
to deformations of the brane it self within the bulk. 

\subsection{The ADD Scenario}\label{subsec:add}

In the ADD\footnote{N. Arkani-Hamed, S. Dimopoulos and G.R. Dvali} scenario it is assumed that 
the extra dimensions are compactified (normally on a torus with equal radii) and large (up to millimeter 
size). This imply that the fundamental scale of gravity would be related to the Planck mass ($M_P$) by
the volume of the large extra dimensions,
\begin{equation}
M_P^2 = 8 \pi R^n M_D^{n+2}
\end{equation}
which would allow $M_D \sim 1$ TeV. Here $n$ corresponds to the number of extra dimensions. These are assumed 
to be torus shaped and R corresponds to the radius of the torus. A second implication
is that the KK-tower would be very dens and could appear as a continuous mass distribution in the experiment, 
since the separation between the mass modes is related to the size of the extra dimensions as 
$\Delta m_{KK} \propto 1/R$. Gravity experiments, cosmology and astrophysical constraints do however put 
bounds on the ADD scenario where the case of one and two extra dimensions are strongly constrained, but for 
more than two extra dimensions the bounds are relatively weak. 

At LEP three different processes were studied in the ADD scenario, within two different categories.
In the first, and most popular, category the brane is assumed to be rigid, $\mbox{f} >> M_D$ where $\mbox{f}$ is 
the brane tension. Signs of extra dimensions could then arise from the two processes 
$e^+e^- \rightarrow \gamma G$ and $e^+e^- \rightarrow G^* \rightarrow ff / \gamma \gamma$. In the process with real 
graviton emission, the cross section is directly sensitive to the number of extra dimensions, $n$, and the fundamental 
scale of gravity, $M_D$, while the other process is sensitive to the ratio 
$\lambda / M_H$. In the later case $M_H$ is an ultraviolet cut-off scale which is not equivalent to $M_D$ 
but should be of the same magnitude and $\lambda$ is a coupling constant which depends on the
underlying theory of gravity, but should be of the order ${\cal O}(1)$ (set equal to $\pm 1$ in the analysis). 
In the second category the brane is assumed to be flexible ($\mbox{f} << M_D$) and the previous two processes
would then be exponentially suppressed. Signs of extra dimensions could still arise from the branon pair 
production process, $e^+e^- \rightarrow \tilde{\pi } \tilde{\pi } + \gamma / Z$ where the signal cross section 
is sensitive to the number of branons $n_b$ ($n_b = 1$ assumed in the analysis), the branon mass $m_b$ and the
brane tension $\mbox{f}$. 

\subsection{The Randall-Sundrum Scenario}\label{subsec:rs}

In the Randall-Sundrum scenario studied at LEP it is assumed that two branes exists. One to which the SM fields
are confined and one where gravity is located, called the Planck brane. It is also assumed that there is only one 
extra dimension and the reason why gravity appear to be weak at the SM brane is due to that it is exponentially 
suppressed with the distance to the Planck brane (due to a ``warped space'' where the normal and extra 
dimensions are not factorisable). Fluctuations in the inter-brane distance give rise to a massive scalar 
called the radion. The radion have the same quantum numbers as the SM Higgs boson which allows them to mix into 
one Higgs-like and one radion-like mass eigenstate. Both these two particles would be produced at LEP mainly through
the Higgsstrahlung process where the final state corresponds to a $Z$ together with either a Higgs or a radion.
This Randall-Sundrum scenario can then be expressed by a mixing parameter $\xi$, the mass scale at the SM brane 
$\Lambda _W$ and the Higgs and radion masses $m_h$ and $m_r$. 

\section{The Graviton Searches}

\begin{figure}[t]
\epsfxsize190pt
\figurebox{190pt}{190pt}{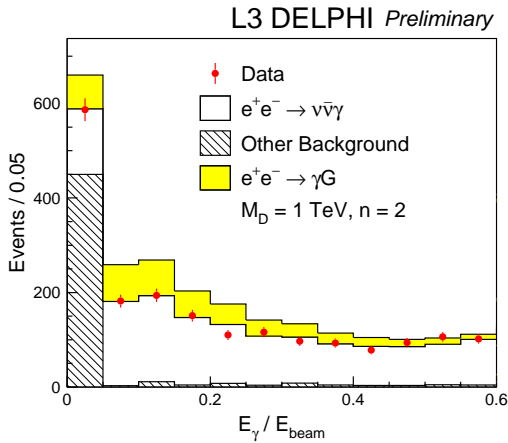}
\caption{Distribution of the ratio of the photon energy to the beam energy 
for single photon events selected by DELPHI and L3, together with the
SM prediction. Expected signal from graviton emission is also shown for 
$M_D = 1$ TeV and $n=2$.}
\label{fig:dldmc}
\end{figure}

\begin{figure}[t]
\epsfxsize180pt
\figurebox{180pt}{180pt}{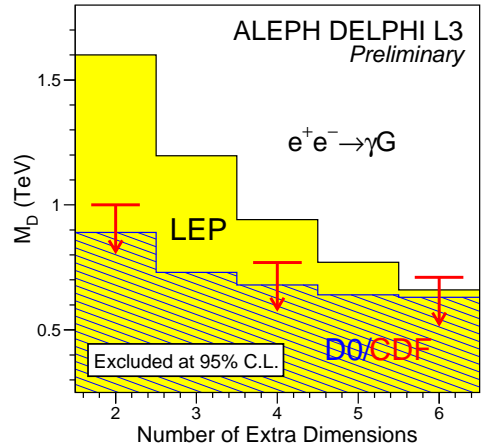}
\caption{The combined ADL $M_D$ limits at 95\% CL as a function of $n$. 
The current limits from the D0 (hatched area) and CDF (indicated by arrows for $n=2$, 4 and 6.) experiments are also shown.}
\label{fig:mdlim}
\end{figure}

The process of graviton production together with a photon is searched for using events where only a single photon is 
detected in the experiment. The distribution of the photon energy (divided by the beam energy) from the combined 
single photon sample selected by DELPHI and L3 is shown in figure \ref{fig:dldmc} together with the predicted graviton 
signal for a value of $M_D = 1$ TeV and $n=2$. 
Since no signs of a signal was observed in any of the LEP experiments, the results were used to derive $M_D$ exclusion 
limits for different numbers of extra dimensions. The results from ALEPH, DELPHI and L3 have also been combined\footnote{OPAL 
was not included in the combination since only results using data recorded up to $\sqrt{s} = 189$ GeV have been reported.} 
and the obtained $M_D$ limits are shown in figure \ref{fig:mdlim} for the number of extra dimensions between two and six. 
The present limits from the D0 and CDF experiments are also shown in the plot.
The obtained $M_D$ limits are summarized in table \ref{tab:mdlim} together with the implied limits on the 
radius of the extra dimensions.

The search for virtual graviton exchange was performed at LEP using many different event topologies where the
highest sensitivity is obtained using Bhabha events, due to the possibility of additional interference with the
Bhabha t-channel diagram. No signs of a signal was observed in any of the experiments and the combined Bhabha
event sample was used to set the strongest limits on $M_H$: $M_H > 1.20$ TeV for $\lambda = +1$
and $M_H > 1.09$ TeV for $\lambda = -1$.

\begin{table}
\caption{The combined ADL $M_D$ limits for the numbers of extra dimensions between 2 and 6.}\label{tab:mdlim}
\begin{center}
\begin{tabular}{|c|c|c|} \hline
\raisebox{0pt}[12pt][6pt]{$n$} &
\raisebox{0pt}[12pt][6pt]{$M_D$ (TeV)} &
\raisebox{0pt}[12pt][6pt]{$R$ (mm)} \\ \hline
\raisebox{0pt}[12pt][6pt]{2} &
\raisebox{0pt}[12pt][6pt]{$> 1.60$} &
\raisebox{0pt}[12pt][6pt]{$< 0.19$} \\\hline
\raisebox{0pt}[12pt][6pt]{3} &
\raisebox{0pt}[12pt][6pt]{$> 1.20$} &
\raisebox{0pt}[12pt][6pt]{$< 2.6 \times 10^{-6}$} \\\hline
\raisebox{0pt}[12pt][6pt]{4} &
\raisebox{0pt}[12pt][6pt]{$> 0.94$} &
\raisebox{0pt}[12pt][6pt]{$< 1.1 \times 10^{-8}$} \\\hline
\raisebox{0pt}[12pt][6pt]{5} &
\raisebox{0pt}[12pt][6pt]{$> 0.77$} &
\raisebox{0pt}[12pt][6pt]{$< 4.1 \times 10^{-10}$} \\\hline
\raisebox{0pt}[12pt][6pt]{6} &
\raisebox{0pt}[12pt][6pt]{$> 0.66$} &
\raisebox{0pt}[12pt][6pt]{$< 4.6 \times 10^{-11}$} \\\hline
\end{tabular}
\end{center}
\end{table}

\section{The TeV String Search}

In addition to the graviton searches, a search for TeV strings was made at the L3 experiment, also using Bhabha events. 
This search looks for possible string effects and not phenomena from graviton production. The string model, however, 
contains ADD extra dimensions which imply that the string scale $M_S$ is related to $M_D$ by a factor of somewhere between 
1.6 to 3. In this model the string effects in the Bhabha scattering process would dominate over the effect from virtual graviton 
exchange and the lack of an observed signal implies an exclusion limit on the string scale of $M_S > 0.55$ TeV.

\section{The Branon Search}

\begin{figure}[t]
\epsfxsize188pt
\figurebox{188pt}{188pt}{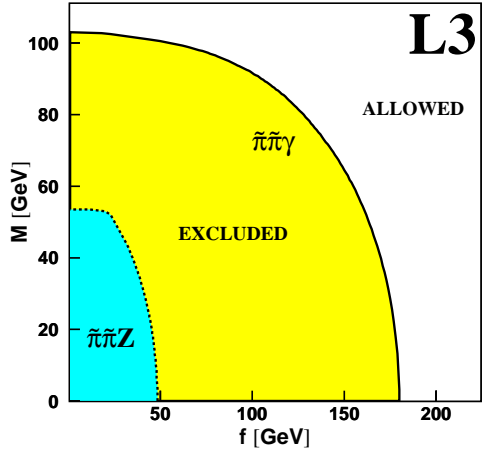}
\caption{The two dimensional regions in the ($\mbox{f}$,$m_b$) plane excluded by the branon searches.}
\label{fig:branon}
\end{figure}

The L3 experiment has also made a search for branons using both single photon events and single $Z$ like hadronic 
events. As for the single photon events the single $Z$ search did not show any signs of a signal and both these 
analysis gave the excluded regions in the $m_b - \mbox{f}$ plane shown in figure \ref{fig:branon}.  
From figure \ref{fig:branon} it is clear that the single photon analysis is the most sensitive one and the
overall limits of $\mbox{f} > 180$ GeV in the case of $m_b = 0$ and $m_b > 103$ GeV in the case of 
$\mbox{f} = 0$ were obtained. 

\section{The Radion Search}

\begin{figure}[t]
\epsfxsize175pt
\figurebox{175pt}{175pt}{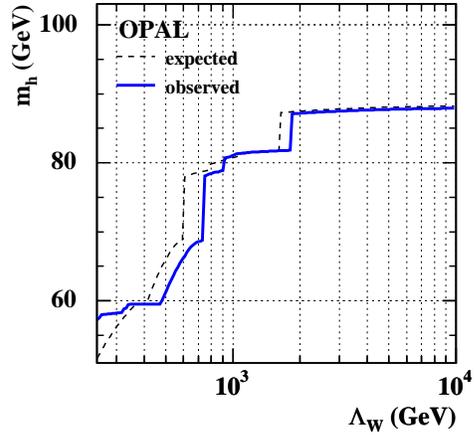}
\caption{The combined ADL $M_D$ limits at 95\% CL as a function of $n$. 
The current limits from the D0 and CDF experiments are also shown.}
\label{fig:radion}
\end{figure}

The results from three different Higgs searches were used in the radion search made at the OPAL experiment.
The three searches corresponds to the standard Higgs search (optimized for $h \rightarrow bb$), a search for a 
hadronically decaying Higgs ($h \rightarrow qq$ or $gg$) and a decay mode independent search where no assumptions 
of the Higgs decay were made. Since no signs of a signal were observed the results were used to constrain the
Randall-Sundrum parameter space, where a point was considered to be excluded if the signal exceeded any of the 
limits from the three analysis used. The search was able to set an exclusion limit on the Higgs mass within the 
whole parameter space. This was, however, not possible for the radion mass, due to the coupling suppression at high values 
of $\Lambda _W$ and a decrease in sensitivity at large negative mixing. The obtained Higgs mass limit is shown 
in figure \ref{fig:radion} as a function only of $\Lambda _W$ and the overall limit of $m_h > 58$ GeV was obtained. 

\end{document}